\documentclass[sigconf]{acmart}
\usepackage{multirow,booktabs}
\usepackage{enumitem}
\usepackage[table]{xcolor}

\definecolor{RowBlue}{HTML}{E6F2FF}
\AtBeginDocument{%
  }

\copyrightyear{2027}
\acmYear{2027}
\setcopyright{cc}
\setcctype{by}
\acmConference[KDD '27]{Proceedings of the 33rd ACM SIGKDD Conference on Knowledge Discovery and Data Mining V.1}{August 01--05, 2027}{San Jose, CA, USA}
\acmBooktitle{Proceedings of the 33rd ACM SIGKDD Conference on Knowledge Discovery and Data Mining V.1 (KDD '27), August 01--05, 2027, San Jose, CA, USA}
\acmDOI{10.1145/3862868.3871513}
\acmISBN{979-8-4007-3134-1/2027/08}

\makeatletter
\patchcmd{\@mkbibcitation}{\authors}
  {Yiqing Yang, Xinlong Zhao, Zhao Liu, Xiao Lv, Ruiming Tang, and Kun Gai}
  {}{\PackageError{DADF}{Could not set the proceedings citation authors}{Check the acmart citation format.}}
\patchcmd{\@mkbibcitation}{\unskip, \ref{TotPages}~\@pages@word}
  {\unskip, 12~pages}
  {}{\PackageError{DADF}{Could not set the proceedings citation length}{Check the acmart citation format.}}
\makeatother

\newcommand{\cofirst}{\authornote{Equal contribution.}}
\newcommand{\cofirstmark}{\authornotemark[1]}
\newcommand{\corr}{\authornote{Corresponding author.}}
\newcommand{\corrmark}{\authornotemark[2]}

\newcommand{\ksaffil}{%
  \affiliation{%
    \institution{Kuaishou Technology}%
    \city{Beijing}%
    \country{China}%
  }%
}

\newcommand{\unaffilcn}{%
  \affiliation{%
    \institution{Unaffiliated}%
    \city{Beijing}%
    \country{China}%
  }%
}

\begin{document}

\title{DADF: A Distribution-Aware Debiasing Framework for Watch-Time Regression in Recommender Systems}

\author{Yiqing Yang}\cofirst
\ksaffil
\email{yangyiqing06@kuaishou.com}

\author{Xinlong Zhao}\cofirstmark
\ksaffil
\email{zhaoxinlong03@kuaishou.com}

\author{Zhao Liu}\cofirstmark
\ksaffil
\email{liuzhao09@kuaishou.com}

\author{Xiaoyou Zhou}
\ksaffil
\email{zhouxiaoyou@kuaishou.com}

\author{Xiao Lv}\corr
\ksaffil
\email{lvxiao03@kuaishou.com}

\author{Ruiming Tang}\corrmark
\ksaffil
\email{tangruiming@kuaishou.com}

\author{Kun Gai}
\unaffilcn
\email{gai.kun@qq.com}

\renewcommand{\shortauthors}{Yiqing Yang et al.}

\begin{abstract}
Watch-time predictors in short-video recommender systems can be well balanced in aggregate and approximately calibrated with respect to their own prediction scores, yet exhibit pronounced mean shrinkage in the label space: when samples are grouped by observed watch time, short observations tend to be overestimated while long observations tend to be underestimated. This ex-post pattern is not itself a calibration violation and may partly arise from irreducible uncertainty. We instead study whether part of it corresponds to conditional residual structure that is predictable from inference-time signals and correctable without replacing a mature first-stage predictor.
We propose DADF, a distribution-aware framework for second-stage multiplicative correction over a frozen watch-time predictor. Here, debiasing refers to correcting the inference-time-predictable component of conditional residuals rather than global recalibration. DADF estimates this predictable correction from the full inference-time representation, stabilizes long-tailed correction targets through group-specific distribution transformations, and incorporates auxiliary representations from engagement prediction heads. Mean shrinkage remains visible after stratifying by video duration, indicating that duration does not explain away the pattern; DADF instead uses duration only to index heterogeneous correction-target distributions and route correction experts.
We evaluate DADF on KuaiRec and WeChat21 with seven first-stage backbones and in a large-scale industrial ranking system. Across offline experiments, DADF reduces MAE by 4.33\% and improves XAUC by 4.01\% on average. In production, it reduces MAE by 12.57\%. Three controlled online A/B tests in full ranking, rough ranking, and degraded serving yield 0.649\%, 0.235\%, and 0.199\% lifts in average time spent per device, respectively, after which all three integrations were deployed to 100\% of production traffic. These results demonstrate that DADF is a practical, model-agnostic plug-in for correcting predictable conditional residuals associated with label-space mean shrinkage without changing the serving interface of mature first-stage models. Code is available at \href{https://github.com/liuzhao09/DADF}{github.com/liuzhao09/DADF}.
\end{abstract}



\begingroup
\renewcommand{\UrlFont}{\normalfont\small}
\maketitle
\endgroup

\begin{figure*}[!t]
  \centering
  \makebox[\textwidth][c]{%
    \includegraphics[height=12cm]{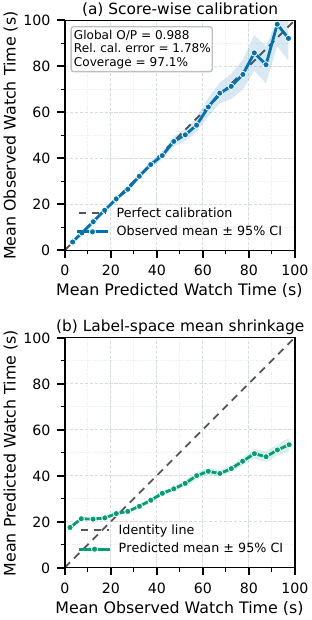}%
    \hspace{0.35cm}%
    \includegraphics[height=12cm]{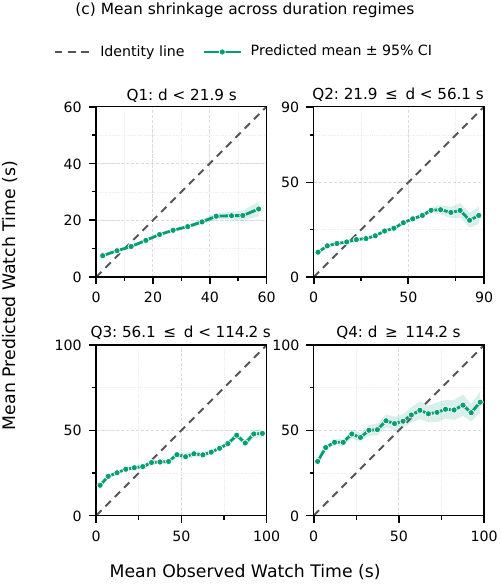}%
  }
  \caption{Motivation for DADF. (a) Prediction-binned observed means closely follow the identity line, indicating approximate score-wise calibration. (b) Without duration stratification, label-binned predictions overestimate short observations and underestimate long ones, revealing label-space mean shrinkage. (c) The same pattern remains visible within every video-duration quartile ($d$ denotes duration in seconds), indicating that coarse duration stratification does not explain it away. Panels (b) and (c) are ex-post label-conditioned diagnostics rather than calibration tests.}
  \label{fig:01_motivation}
\end{figure*}

\section{Introduction}

Watch time is a central continuous target in short-video recommender
systems because it captures consumption depth more precisely than
binary feedback such as clicks or likes. Unlike binary actions that
mainly indicate whether an interaction occurred, watch time further
distinguishes brief skips, effective consumption, and sustained
engagement. Prediction errors in different consumption regions can
alter candidate ordering: systematically overestimated content may
receive excessive exposure, whereas underestimated content associated
with deep consumption may be suppressed. Such local errors may
partially cancel in aggregate statistics while still distorting
exposure allocation and traffic distribution.

However, mature first-stage predictors are tightly coupled with feature
pipelines and serving infrastructure, making replacement costly. A
production watch-time predictor is typically integrated with feature
generation, multi-task training, model refresh, candidate scoring, and
downstream ranking objectives. Replacing it may require feature
migration, capacity evaluation, interface adaptation, and coordinated
changes across multiple serving stages. A lightweight second-stage
module that freezes the existing predictor and preserves the scalar
watch-time interface therefore provides a lower-risk path for
incremental improvement. We study this minimally intrusive correction
setting using only inference-time signals.

Let $Y$ denote observed watch time and $\hat{Y}_0$ the frozen
prediction. Figure~\ref{fig:01_motivation}(a) shows that
prediction-binned observed means closely follow the identity line, so
the predictor is approximately calibrated with respect to its own
scores. In contrast, conditioning on the realized label reveals
predictions above the identity line for short observations and below it
for long observations
(Figure~\ref{fig:01_motivation}(b)). We call this compression toward
the population mean \emph{label-space mean shrinkage}.

These two diagnostics reverse the conditioning direction. Score-wise
calibration asks whether the prediction can be interpreted at face
value among samples receiving that score, whereas the label-space view
examines predictions among samples sharing a similar realized outcome.
A model can therefore be approximately calibrated with respect to its
own scores and still exhibit pronounced label-space shrinkage. Because
the latter is an ex-post label-conditioned diagnostic, it is not itself
evidence of a calibration violation and may partly reflect irreducible
uncertainty.

Consequently, we do not attempt to mechanically invert the entire curve
in Figure~\ref{fig:01_motivation}(b), nor assume that all observed
shrinkage is removable. We instead ask whether the frozen prediction
leaves conditional residual structure that remains predictable from
information available at serving time. Only this
inference-time-predictable component is considered correctable.

Figure~\ref{fig:01_motivation}(c) repeats the diagnostic within
video-duration quartiles. Mean shrinkage remains visible in every
regime, indicating that coarse duration stratification does not explain
away the pattern. This observation also clarifies the role of duration
in our framework. The pattern cannot be attributed simply to a mixture
of short and long videos; nevertheless, duration remains a stable
inference-time variable that can index regions with different residual
statistics without being interpreted as the causal source of those
residuals.

Meanwhile, the multiplicative correction target
$B=Y/\max(\hat{Y}_0,\epsilon)$ exhibits regime-dependent skewness,
dispersion, and tail behavior
(Figure~\ref{fig:07_factor_distribution_compare}). The ratio form is
attractive because it expresses correction relative to the deployed
prediction and preserves the scalar interface after multiplication. It
is also challenging to optimize: watch time is long-tailed, and
division by a small first-stage prediction can further amplify the
target range. Moreover, correction factors from different duration
regimes may have substantially different scales, skewness, and tail
geometry. These differences motivate using duration only as an
observable index of residual heterogeneity, while the correction itself
must be inferred from richer inference-time representations.

Prior work mainly replaces or retrains the first-stage predictor through
discretization~\cite{tpm,cread}, duration-aware
modeling~\cite{d2q,d2co}, or distributional modeling~\cite{egmn}.
These approaches improve the primary estimation problem by redesigning
its learning target, output structure, or conditional distribution
model. We study a complementary industrial setting in which the primary
predictor is already mature, deployed, and costly to replace, and the
goal is to correct residual structure without modifying its training
objective or serving interface.

We adapt TranSUN's jointly trained multiplicative correction
mechanism~\cite{transsun} to our frozen-backbone protocol as a
controlled MLP baseline. Across two datasets and seven backbones, this
adaptation improves held-out MAE in 10 of 14 settings, showing that even
a simple correction branch can recover part of the predictable
residual. However, the adapted baseline does not explicitly address
three finite-model challenges: the long-tailed and skewed ratio-based
correction target, its distributional heterogeneity across duration
regimes, and the incremental residual-predictive information available
in auxiliary engagement signals. Directly fitting the raw ratio allows
extreme samples to dominate optimization; a shared correction mapping
may be insufficient for heterogeneous regimes; and a scalar watch-time
output may omit behavioral states already represented by related
engagement heads.

We propose \textbf{DADF}, a \textbf{D}istribution-\textbf{A}ware
\textbf{D}ebiasing \textbf{F}ramework for second-stage watch-time
correction. Here, debiasing refers specifically to correcting
predictable conditional residuals rather than globally recalibrating or
replacing the first-stage model. DADF addresses the three challenges
through group-specific distribution transformation, duration-indexed
expert routing, and multi-label-aware auxiliary representation
learning.

Specifically, DADF learns a separate monotone target transformation for
each duration regime, mapping the long-tailed correction factor into a
more stable regression space. It then applies deterministic hard
routing so that each impression activates only one correction expert,
allowing finite-capacity experts to specialize without serving-time
multi-expert fan-out. Finally, DADF reuses logits and tower
representations from related engagement objectives. Their fixed
nonlinear responses are fused with the shared correction context to
provide behavioral states that may not be fully expressed by the scalar
watch-time prediction.

During training, observed watch time and auxiliary labels are used only
to construct supervision, while the first-stage prediction is frozen
through stop-gradient. At serving time, neither observed watch time,
realized residuals, nor auxiliary labels are available. The correction
network uses only the frozen prediction and inference-time
representations to produce a positive multiplicative factor, which is
multiplied by the first-stage output. DADF therefore retains the
first-stage model and serving interface as a minimally intrusive,
model-agnostic plug-in.

Our main contributions are:
\begin{itemize}[leftmargin=15pt, topsep=2pt, itemsep=1pt,
parsep=0pt, partopsep=0pt]
    \item We distinguish score-wise calibration from label-space mean
    shrinkage and formulate second-stage correction of its
    inference-time-predictable component over a frozen watch-time
    predictor, without assuming that all observed shrinkage is
    removable.

    \item We propose DADF to address the heavy-tailed ratio-based
    correction target, duration-indexed distributional heterogeneity,
    and auxiliary-signal utilization through three corresponding
    modules, while preserving the existing scalar prediction interface.

    \item We demonstrate accurate and robust correction across public
    benchmarks and three production ranking stages, followed by
    deployment to 100\% of traffic without changing the existing
    serving interface.
\end{itemize}
\section{Related Work}

\subsection{Watch-Time Prediction and Duration-Aware Modeling}

Watch time and dwell time are widely used as continuous engagement signals in recommender systems~\cite{covington2016youtube,yi2014beyondclicks,wu2018beyondviews}. Recent watch-time predictors improve the primary estimation problem through ordinal decomposition and discretization~\cite{tpm,cread}, conditional quantile estimation~\cite{cqe}, or explicit distribution modeling~\cite{wang2020capturing,egmn}. These approaches address the long-tailedness, uncertainty, or multimodality of the watch-time label by redesigning and training the primary predictor.

A related line explicitly models how video duration affects exposure, observed watch time, or the interpretation of user interest. D2Q applies causal adjustment to duration-related exposure effects~\cite{d2q}; DVR optimizes watch-time gain under duration bias~\cite{dvr}; D$^2$CO separates duration-related and noisy components from observed watch time~\cite{d2co}; and CWM estimates counterfactual watch time under duration truncation~\cite{cwm}. These methods modify the primary learning target or predictor to address a duration-related observation mechanism. DADF studies a different setting: it neither attributes the residual error to duration nor estimates duration-neutral interest. It retains a frozen watch-time predictor and uses duration as an observable index for heterogeneous second-stage correction-target distributions.

\subsection{Regression Diagnostics and Multiplicative Correction}

Calibration theory for stand-alone point forecasts asks whether a prediction can be taken at face value conditional on the issued score, and evaluates this property through reliability diagrams~\cite{gneiting2023regression}. This prediction-conditioned notion is distinct from the label-conditioned mean-shrinkage diagnostic in Figure~\ref{fig:01_motivation}. Separately, transforming a skewed continuous target can produce a more regular optimization space, as in Box--Cox-style regression~\cite{sakia1992boxcox}. Because a nonlinear inverse map does not generally preserve conditional expectations, transformed regression can incur retransformation error; classical remedies include smearing estimation and related statistical corrections~\cite{duan1983smearing,newman1993log,stow2006bayesian}. TranSUN addresses this issue within recommender regression by introducing a jointly trained multiplicative correction branch and generalizing it into a family of transformation-aware regression models~\cite{transsun}.

DADF adopts the deployable form of multiplicative correction but targets a different problem: predictable conditional residual structure left by an arbitrary frozen watch-time predictor. For controlled evaluation, we adapt the TranSUN correction mechanism to the same frozen-backbone protocol used by DADF. Beyond this common correction form, DADF explicitly models the heavy-tailed and skewed ratio target, its distributional heterogeneity across duration-indexed regimes, and the auxiliary engagement representations available at inference time. The contribution is therefore not target transformation alone, but its integration with regime-specific correction and inference-time side information under a minimally intrusive second-stage setting.

\subsection{Auxiliary Representations in Multi-Task Recommendation}

Multi-task recommenders share information across related feedback objectives through architectures such as MMoE~\cite{mmoe}, PLE~\cite{ple}, ESMM~\cite{esmm}, and AITM~\cite{aitm}. Their auxiliary supervision is typically used to improve jointly trained task predictors. DADF instead reuses the logits and tower representations of engagement heads as inference-time conditioning signals for correction over a frozen watch-time predictor. Observed auxiliary labels supervise the corresponding heads during training but are never inputs to the correction model during serving. This design uses behavioral representations to reduce conditional uncertainty in the predicted correction without changing the first-stage prediction interface.

\section{Problem Formulation}
\label{sec:problem}

\subsection{Frozen-Predictor Correction Setting}

For each impression, let $Y\in\mathbb{R}_{+}$ denote observed watch time and let $\hat{Y}_0\in\mathbb{R}_{+}$ be the output of a fixed first-stage predictor. We consider a minimally intrusive second stage that leaves this predictor and its serving interface unchanged. Let $R=(\hat{Y}_0,x,\ell_0,a)$ collect the correction representation: ranking features $x$, the base prediction $\hat{Y}_0$, its raw output $\ell_0$ before the final output mapping, and auxiliary engagement representations $a$. Video duration $d$ is mapped to a pre-defined regime $G=\pi(d)$, and all information available to the correction model at serving time is denoted by $S=(R,G)$. The observed watch time and auxiliary labels may supervise training but are unavailable as correction inputs during inference.

\subsection{From Label-Space Diagnosis to Correctable Residuals}

Mean calibration of a point predictor is prediction-conditioned and requires $\mathbb{E}[Y\mid\hat{Y}_0]=\hat{Y}_0$~\cite{gneiting2023regression}. By contrast, Figure~\ref{fig:01_motivation}(b) examines $\mathbb{E}[\hat{Y}_0\mid Y]$ and reveals label-space mean shrinkage. Because it conditions on the realized label, this curve is an ex-post error diagnostic rather than evidence of a calibration violation. It can also contain irreducible variation that no inference-time model can recover.

We therefore define the correctable additive residual as
\begin{equation}
  r^{*}(S)=\mathbb{E}[Y-\hat{Y}_0\mid S],
\end{equation}
which yields the population-optimal second-stage predictor $f^{*}(S)=\hat{Y}_0+r^{*}(S)=\mathbb{E}[Y\mid S]$. Under squared loss, since $\hat{Y}_0$ is measurable with respect to $S$,
\begin{equation}
  \mathcal{R}(\hat{Y}_0)-\mathcal{R}(f^{*})
  =
  \mathbb{E}\!\left[
    \left(\mathbb{E}[Y\mid S]-\hat{Y}_0\right)^2
  \right]\ge 0.
  \label{eq:correctable_risk}
\end{equation}
 The inequality is strict only when $S$ contains conditional-mean information not already expressed by $\hat{Y}_0$(see Appendix~\ref{app:predictable_risk_proof}). This identity characterizes the population opportunity under squared loss; it neither guarantees improvement for a finite correction model nor implies that all label-space shrinkage is removable. Whether the residual is predictable, and whether correction improves MAE and ranking utility, must be established on held-out and online data.

\subsection{Multiplicative Correction Target}

Following the multiplicative correction paradigm, we define the training target
\begin{equation}
  B=\frac{Y}{\max(\operatorname{sg}(\hat{Y}_0),\epsilon)},
  \label{eq:correction_target}
\end{equation}
where $\operatorname{sg}(\cdot)$ denotes stop-gradient and $\epsilon>0$ prevents numerical instability near zero. A correction network predicts $\hat{B}=c_{\theta}(S)>0$ and returns
\begin{equation}
  \hat{Y}=\hat{Y}_0\hat{B}.
  \label{eq:corrected_prediction}
\end{equation}
Because the denominator in Eq.~\eqref{eq:correction_target} is determined by $S$,
\begin{equation}
  \mathbb{E}[B\mid S]
  =
  \frac{\mathbb{E}[Y\mid S]}
       {\max(\hat{Y}_0,\epsilon)}.
\end{equation}
Thus, for $\hat{Y}_0\ge\epsilon$, the population-optimal factor recovers $\mathbb{E}[Y\mid S]$ after multiplication. For smaller predictions, clipping trades exact recovery for numerical stability. Importantly, $B$ is a noisy sample-level correction target rather than a ground-truth bias factor. DADF estimates its predictable conditional structure; duration only indexes the transformation and expert associated with $G$, and no observed watch time or realized residual is used at serving time.

\section{Method}
\label{sec:method}

\begin{figure}[t]
  \centering
  \includegraphics[width=\linewidth]{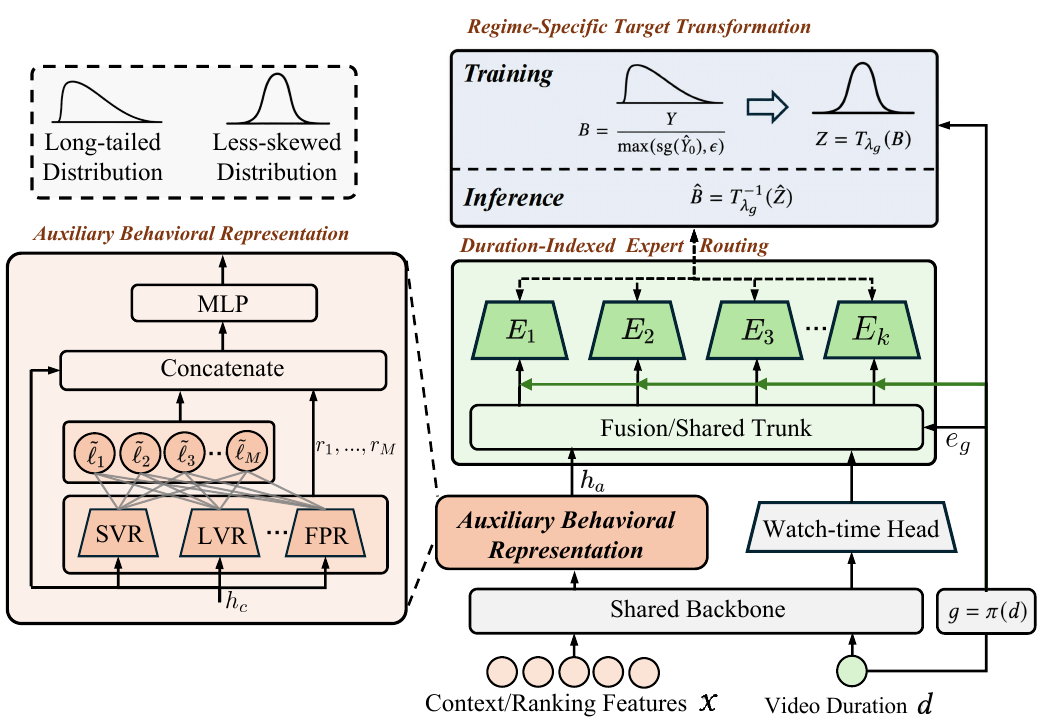}
  \caption{Overview of DADF. Training constructs a multiplicative correction target from the frozen first-stage prediction and observed watch time. Serving uses only inference-time representations and the duration-indexed regime to predict and invert the correction.}
  \label{fig:02_framework}
\end{figure}

DADF estimates the predictable multiplicative correction defined in Section~\ref{sec:problem} while keeping the first-stage watch-time predictor fixed. Its three components correspond to three finite-model challenges: a skewed and heavy-tailed correction target, heterogeneous target distributions across duration-indexed regimes, and residual-predictive information in auxiliary engagement representations. As shown in Figure~\ref{fig:02_framework}, DADF transforms the target within each regime, predicts it with a duration-routed expert conditioned on the full correction representation, and maps the output back to the original watch-time space.

\subsection{Regime-Specific Target Transformation}
\label{sec:dynamic_distribution}

The sample target $B$ in Eq.~\eqref{eq:correction_target} inherits the long-tailedness of watch time under mild conditions (Appendix~\ref{app:long_tail_factor}), and division by a small base prediction can further amplify its dispersion. Directly fitting this target allows a small number of extreme samples to dominate optimization. Moreover, the empirical skewness, dispersion, and tail behavior of $B$ vary across duration regimes (Figure~\ref{fig:07_factor_distribution_compare}). These marginal differences motivate regime-specific target stabilization; they are not interpreted as evidence that duration causes the residual error.

For a sample with duration $d$, let $g=\pi(d)\in\mathcal{G}$ denote its pre-defined regime. DADF applies a group-specific Box--Cox-style map:
\begin{equation}
  T_{\lambda_g}(B)=
  \begin{cases}
    \dfrac{\max(B,\epsilon)^{\lambda_g}-1}{\lambda_g},
      & \lambda_g\neq 0,\\[5pt]
    \log\max(B,\epsilon), & \lambda_g=0,
  \end{cases}
  \label{eq:group_transform}
\end{equation}
where $\lambda_g$ is learned jointly with the correction network. The map is monotone for positive targets, so it changes target geometry without changing within-regime order. A separate $\lambda_g$ allows each regime to use a transformation strength suited to its correction-target distribution instead of imposing one global transform.

Let $Z=T_{\lambda_g}(B)$ be the transformed target. The corresponding inverse is
\begin{equation}
  T_{\lambda_g}^{-1}(z)=
  \begin{cases}
    (\lambda_g z+1)^{1/\lambda_g}, & \lambda_g\neq 0,\\
    \exp(z), & \lambda_g=0.
  \end{cases}
  \label{eq:group_inverse}
\end{equation}
The implementation clips the predicted transformed value to the valid numerical range before applying Eq.~\eqref{eq:group_inverse}; detailed bounds are reported in Section~\ref{sec:experiments}. Regime assignments and learned transformation parameters are fixed after training, and both are available without observing the realized watch time.

\subsection{Duration-Indexed Expert Routing}
\label{sec:debias_factor}

The correction mapping may also differ across regimes under finite model capacity. DADF forms a shared expert input
\begin{equation}
  h=\operatorname{Fusion}(\hat{Y}_0,x,\ell_0,h_a,e_g),
  \label{eq:correction_fusion}
\end{equation}
where $e_g$ is the embedding of regime $g$ and $h_a$ is the auxiliary behavioral representation introduced below. It then uses deterministic hard routing:
\begin{equation}
  \hat{Z}=E_g(h),
  \label{eq:hard_routing}
\end{equation}
where $E_g$ is the expert assigned to regime $g$. The expert consumes the full correction representation rather than predicting from duration alone. Hard routing activates one expert per impression, avoids serving-time fan-out, and lets finite-capacity experts specialize to regime-dependent correction mappings.

This routing is an architectural use of duration, not a causal claim about the origin of mean shrinkage. Distinct marginal target distributions motivate specialization, while its practical contribution is evaluated through ablation rather than assumed from the distribution plot. Because routing also changes parameter specialization, the ablation is not interpreted as isolating duration from every matched-capacity alternative. The released default and all main results use hard routing; soft routing is retained only as an ablation.

 We further discuss the representational benefit and limitations of duration-indexed hard routing in Appendix~\ref{app:group_risk_proof}.
\subsection{Auxiliary Behavioral Representation}

Related engagement objectives can reveal behavioral states that are not fully expressed by the scalar watch-time prediction. For each auxiliary task $m$, a task-specific tower and head produce
\begin{equation}
  r_m=T_m(h_c), \qquad
  \ell_m=H_m(r_m), \qquad m=1,\ldots,M,
\end{equation}
where $h_c$ is the shared correction context, $r_m$ is the tower representation, and $\ell_m$ is the auxiliary logit. DADF expands each logit through fixed nonlinear responses:
\begin{equation}
  \begin{aligned}
  \tilde{\ell}_m
  =
  \Big[&
  \ell_m,\,
  \operatorname{ReLU}(\ell_m),\,
  \sigma(\ell_m),\,
  \frac{\tanh(\ell_m)+1}{2},\,
  \ell_m^2,\\
  &
  \sqrt{\operatorname{ReLU}(\ell_m)+\epsilon},\,
  \log\!\bigl(1+\operatorname{ReLU}(\ell_m)\bigr),\,
  \operatorname{softplus}(\ell_m)
  \Big],
  \end{aligned}
  \label{eq:aux_logit_expansion}
\end{equation}
and fuses logits and tower states as
\begin{equation}
  h_a
  =
  \operatorname{MLP}\!\left(
    \left[
      h_c,\,
      \tilde{\ell}_1,\ldots,\tilde{\ell}_M,\,
      r_1,\ldots,r_M
    \right]
  \right).
  \label{eq:aux_representation}
\end{equation}
The copies entering the correction branch use stop-gradient, so auxiliary heads are optimized by their own supervision rather than through the correction loss. In production, DADF reuses logits and tower states from existing first-stage engagement heads; the public implementation trains lightweight counterparts from available labels. In both cases, observed auxiliary labels supervise training only and are never serving-time inputs.

\subsection{Training Objective and Inference}

For a sample in regime $g$, the correction network predicts
\begin{equation}
  \hat{Z}=F_{\theta}(\hat{Y}_0,x,\ell_0,g,h_a),
  \qquad
  \hat{B}=T_{\lambda_g}^{-1}(\hat{Z}),
  \qquad
  \hat{Y}=\hat{Y}_0\hat{B}.
\end{equation}
The transformed-space regression loss is
\begin{equation}
  \mathcal{L}_{\mathrm{trans}}=\ell(Z,\hat{Z}),
\end{equation}
where $Z=T_{\lambda_g}(B)$ and $\ell$ is MSE in the reported implementation. To constrain the quantity consumed by the downstream ranker, DADF also applies
\begin{equation}
  \mathcal{L}_{\mathrm{abs}}
  =
  \ell_{\mathrm{Huber}}(Y,\hat{Y})
\end{equation}
in the original watch-time space.

To stabilize the learned transformation, let $\mu_g$, $\sigma_g^2$, and $\operatorname{Skew}_g$ be the mini-batch moments of $Z$ for regime $g$. DADF uses
\begin{equation}
  \mathcal{L}_{\mathrm{reg}}
  =
  \frac{1}{|\mathcal{G}|}
  \sum_{g\in\mathcal{G}}
  \left[
    w_{\mu}\mu_g^2+
    w_{\sigma}(\sigma_g^2-1)^2+
    w_s|\operatorname{Skew}_g|
  \right],
\end{equation}
masking regimes with insufficient mini-batch support. When lightweight auxiliary heads are trained, their supervision is
\begin{equation}
  \mathcal{L}_{\mathrm{aux}}
  =
  \frac{1}{M}\sum_{m=1}^{M}
  \operatorname{BCE}(q_m,\sigma(\ell_m)),
\end{equation}
where $q_m$ is used only as a training label. The complete objective is
\begin{equation}
  \mathcal{L}_{\mathrm{DADF}}
  =
  \alpha\mathcal{L}_{\mathrm{trans}}
  +\beta\mathcal{L}_{\mathrm{abs}}
  +\eta\mathcal{L}_{\mathrm{reg}}
  +\gamma\mathcal{L}_{\mathrm{aux}}.
  \label{eq:dadf_loss}
\end{equation}

The original-space term anchors the jointly learned transformation to the downstream prediction objective, while the transformed-space and moment terms improve optimization geometry; the transformation is therefore not optimized solely to normalize batch statistics.

At serving time, the observed watch time, realized correction target, and auxiliary labels are unavailable. Given the frozen prediction and inference-time signals, DADF returns
\begin{equation}
  \hat{Y}
  =
  \hat{Y}_0\,
  T_{\lambda_{\pi(d)}}^{-1}
  \!\left(F_{\theta}(\hat{Y}_0,x,\ell_0,\pi(d),h_a)\right).
\end{equation}
Thus, the full inference-time representation predicts the correction, while duration indexes the fixed transformation and expert used for that prediction.

\section{Experiments}
\label{sec:experiments}

We evaluate DADF from four complementary perspectives:
\begin{itemize}[leftmargin=*,nosep]
  \item \textbf{RQ1 (Correctability and effectiveness):} Do frozen watch-time predictors retain residual structure that a second stage can exploit, and does DADF improve diverse backbones?
  \item \textbf{RQ2 (Component contribution):} How do target transformation, duration-indexed routing, and auxiliary representations contribute to performance?
  \item \textbf{RQ3 (Fine-grained behavior):} Do the gains persist across duration and observed-watch-time regions, long-duration tails, and alternative regime granularities?
  \item \textbf{RQ4 (Industrial utility):} Does DADF improve production accuracy and online engagement while preserving the deployed prediction interface?
\end{itemize}

\subsection{Experimental Setup}

\subsubsection{Datasets}

We use two public short-video datasets with watch-time feedback. \textbf{KuaiRec}\footnote{\url{https://kuairec.com/}}~\cite{kuaiRec} contains 12,530,806 impressions from 7,176 users and 10,728 videos; its dense user--video interactions reduce exposure-selection sparsity relative to ordinary logs. \textbf{WeChat21}\footnote{\url{https://algo.weixin.qq.com/}} contains 7,310,108 interactions between 20,000 users and 96,418 videos, providing a larger and sparser complementary benchmark.

\subsubsection{Baselines}

We evaluate seven first-stage backbones spanning direct regression, discretized estimation, duration-aware modeling, and distributional prediction:
\begin{itemize}[leftmargin=*,nosep]
  \item \textbf{VR}: direct value regression with a pointwise loss.
  \item \textbf{WLR}~\cite{covington2016youtube}: weighted logistic regression for expected watch time.
  \item \textbf{TPM}~\cite{tpm}: ordinal watch-time decomposition through a progressive tree.
  \item \textbf{D2Q}~\cite{d2q}: duration-deconfounded quantile-based watch-time prediction.
  \item \textbf{CREAD}~\cite{cread}: error-adaptive discretization followed by value restoration.
  \item \textbf{D$^2$CO}~\cite{d2co}: correction of duration-related and noisy components when estimating user interest.
  \item \textbf{EGMN}~\cite{egmn}: exponential--Gaussian mixture modeling of watch-time distributions.
\end{itemize}
For each backbone, we compare the uncorrected model, an adaptation of the TranSUN multiplicative correction mechanism~\cite{transsun}, and DADF. Both correction methods use the same frozen first-stage checkpoint. Because the original TranSUN jointly trains its correction branch with transformed regression, our implementation adapts that branch to the frozen-backbone protocol required by this study; it should therefore be interpreted as a controlled multiplicative-correction baseline rather than an exact reproduction of TranSUN's original training procedure.

\subsubsection{Metrics}

Following prior watch-time studies~\cite{tpm,d2q,cread,egmn}, we report mean absolute error (MAE) and pairwise ranking consistency (XAUC):
\begin{equation}
  \mathrm{MAE}
  =
  \frac{1}{N}\sum_{i=1}^{N}|\hat{y}_i-y_i|,
\end{equation}
\begin{equation}
  \mathrm{XAUC}
  =
  \frac{1}{|\Omega|}
  \sum_{(i,j)\in\Omega}
  \mathbf{1}\!\left[
    (\hat{y}_i-\hat{y}_j)(y_i-y_j)>0
  \right],
\end{equation}
where $\Omega=\{(i,j)\mid 1\le i<j\le N\}$. Lower MAE and higher XAUC are better; label or prediction ties receive zero XAUC credit. These metrics evaluate regression and ranking quality, not score-wise calibration.

\subsubsection{Implementation Details}

All methods use an 8:1:1 split with identical preprocessing and raw features; DADF adds no external data. Backbones use 16-dimensional embeddings and 256--128--64 MLPs. Correction methods share frozen checkpoints and seeds. Across seven backbones and 10 seeds, matching each backbone to Backbone+DADF dense capacity within 0.1\% increased dense parameters by 31.79\% on average but reduced XAUC by 0.0013. Offline DADF uses equal-frequency duration regimes ($K=4$ on KuaiRec and $K=3$ on WeChat21); production uses four fixed regimes selected from historical traffic statistics and engineering experience. We set $(\alpha,\beta,\eta,\gamma)=(1.0,0.8,0.05,0.10)$ in Eq.~\eqref{eq:dadf_loss} and select remaining hyperparameters on validation data.

The reported default uses hard routing, group-level $\lambda_g$, and uniform group weighting. Transformed and absolute losses mask $B\notin[0.001,100]$ or $Z\notin[-4,4]$, while moment regularization uses $Z\in[-6,6]$. The inverse path clips $\hat Z$ to $[-6,6]$ and the recovered factor to $[0.1,10]$. For each metric, DADF is compared with the strongest baseline in its backbone group using a paired $t$-test over matched-seed runs ($p<0.05$). Detailed capacity-control results and reproduction commands are provided in the repository.

\subsection{Residual Correctability and Offline Effectiveness (RQ1)}

Table~\ref{tab:01_main_results} reports results on both datasets. The adapted TranSUN correction improves MAE over its frozen backbone in 10 of 14 backbone--dataset settings. This is not universal, but it provides held-out evidence that a simple multiplicative branch can recover part of the residual structure for many first-stage predictors.

DADF achieves the best MAE and XAUC in all 14 settings. Relative to the corresponding uncorrected backbones, it reduces MAE by 4.33\% and improves XAUC by 4.01\% on average. It also outperforms the adapted TranSUN baseline in every backbone group, with each metric-specific gain over the strongest baseline significant under the matched-seed paired test. The result supports two bounded conclusions: residual correction is useful beyond a single backbone family, and DADF is more robust than the simpler multiplicative baseline under the common frozen-backbone protocol.

\begin{table}[t]
\centering
\caption{
Overall performance on KuaiRec and WeChat21 by MAE$\downarrow$ and XAUC$\uparrow$.
Best results within each backbone group are in \textbf{bold}, and second-best are \underline{underlined}.
For each metric, DADF significantly outperforms the strongest baseline in the corresponding backbone group under a paired $t$-test over matched-seed runs ($p<0.05$).
}
\label{tab:01_main_results}
\setlength{\tabcolsep}{3.0pt}
\renewcommand{\arraystretch}{1.05}
\begin{tabular}{l|crrrr}
\toprule
\multirow{2}{*}{Backbone} & \multirow{2}{*}{Method}
  & \multicolumn{2}{c}{KuaiRec}
  & \multicolumn{2}{c}{WeChat21} \\
\cmidrule(lr){3-4}\cmidrule(l){5-6}
  & & MAE$\downarrow$ & XAUC$\uparrow$ & MAE$\downarrow$ & XAUC$\uparrow$ \\
\midrule
\multirow{3}{*}{\textit{VR}}
  & Base & 4.584 & 0.5578 & 18.681 & 0.6766 \\
  & w/ TranSUN & \underline{4.478} & \underline{0.5693} & \underline{18.571} & \underline{0.6787} \\
  & \textbf{w/ DADF} & \textbf{4.235} & \textbf{0.6125} & \textbf{17.912} & \textbf{0.6902} \\
\midrule
\multirow{3}{*}{\textit{WLR}}
  & Base & 4.414 & 0.5941 & 18.215 & 0.6861 \\
  & w/ TranSUN & \underline{4.364} & \underline{0.5965} & \underline{18.133} & \underline{0.6876} \\
  & \textbf{w/ DADF} & \textbf{4.172} & \textbf{0.6227} & \textbf{17.838} & \textbf{0.6934} \\
\midrule
\multirow{3}{*}{\textit{TPM}}
  & Base & 4.459 & 0.5495 & 19.545 & 0.6570 \\
  & w/ TranSUN & \underline{4.361} & \underline{0.5971} & \underline{18.529} & \underline{0.6814} \\
  & \textbf{w/ DADF} & \textbf{4.166} & \textbf{0.6233} & \textbf{18.109} & \textbf{0.6898} \\
\midrule
\multirow{3}{*}{\textit{D2Q}}
  & Base & \underline{4.123} & \underline{0.6319} & \underline{17.544} & \underline{0.6935} \\
  & w/ TranSUN & 4.323 & 0.6082 & 17.855 & 0.6925 \\
  & \textbf{w/ DADF} & \textbf{4.106} & \textbf{0.6345} & \textbf{17.534} & \textbf{0.6946} \\
\midrule
\multirow{3}{*}{\textit{CREAD}}
  & Base & \underline{4.346} & 0.5927 & 19.128 & 0.6679 \\
  & w/ TranSUN & 4.395 & \underline{0.5958} & \underline{18.515} & \underline{0.6824} \\
  & \textbf{w/ DADF} & \textbf{4.189} & \textbf{0.6211} & \textbf{18.164} & \textbf{0.6903} \\
\midrule
\multirow{3}{*}{\textit{D$^2$CO}}
  & Base & 4.613 & 0.5687 & 18.558 & 0.6861 \\
  & w/ TranSUN & \underline{4.300} & \underline{0.6097} & \underline{18.080} & \underline{0.6868} \\
  & \textbf{w/ DADF} & \textbf{4.168} & \textbf{0.6233} & \textbf{17.683} & \textbf{0.6952} \\
\midrule
\multirow{3}{*}{\textit{EGMN}}
  & Base & \underline{4.081} & \underline{0.6245} & 18.330 & \underline{0.6896} \\
  & w/ TranSUN & 4.255 & 0.6120 & \underline{18.099} & 0.6892 \\
  & \textbf{w/ DADF} & \textbf{4.002} & \textbf{0.6257} & \textbf{17.955} & \textbf{0.6911} \\
\bottomrule
\end{tabular}
\end{table}

\subsection{Component Contribution (RQ2)}

We evaluate three variants on the WLR backbone. \textbf{w/o Dist.} removes regime-specific target transformation and fits the raw correction target. \textbf{w/o Factor}, the legacy table label for removing duration-indexed regime routing, replaces the routed correction mapping with its shared counterpart. \textbf{w/o Aux.} removes auxiliary engagement logits and tower representations.

As shown in Table~\ref{tab:03_ablation}, every removal degrades MAE and XAUC on both datasets. Target transformation gives consistent improvements, while auxiliary representations produce the largest XAUC decrease when removed and the largest MAE decrease on WeChat21. Removing regime routing has a smaller but consistent effect. These results show that the three components are complementary within the complete system. They do not by themselves establish a causal role for duration or rule out every matched-input, matched-capacity alternative.

\begin{table}[!ht]
\centering
\caption{Ablation study results of DADF on WLR~\cite{covington2016youtube}.}
\label{tab:03_ablation}
\setlength{\tabcolsep}{5pt}
\renewcommand{\arraystretch}{1.05}
\begin{tabular}{l|rrrr}
\toprule
Variant
  & \multicolumn{2}{c}{KuaiRec}
  & \multicolumn{2}{c}{WeChat21} \\
\cmidrule(lr){2-3}\cmidrule(l){4-5}
  & MAE$\downarrow$ & XAUC$\uparrow$ & MAE$\downarrow$ & XAUC$\uparrow$ \\
\midrule
Full DADF & 4.1723 & 0.6227 & 17.8376 & 0.6934 \\
\midrule
w/o Dist. & 4.1901 & 0.6210 & 17.8748 & 0.6930 \\
w/o Factor & 4.1823 & 0.6212 & 17.8454 & 0.6931 \\
w/o Aux. & 4.1865 & 0.6204 & 17.9137 & 0.6920 \\
\bottomrule
\end{tabular}
\end{table}

\subsection{Fine-Grained Error Analysis (RQ3)}

\subsubsection{Duration and Label-Space Regions}

\begin{figure}[htbp]
    \centering
    \includegraphics[width=0.85\linewidth]{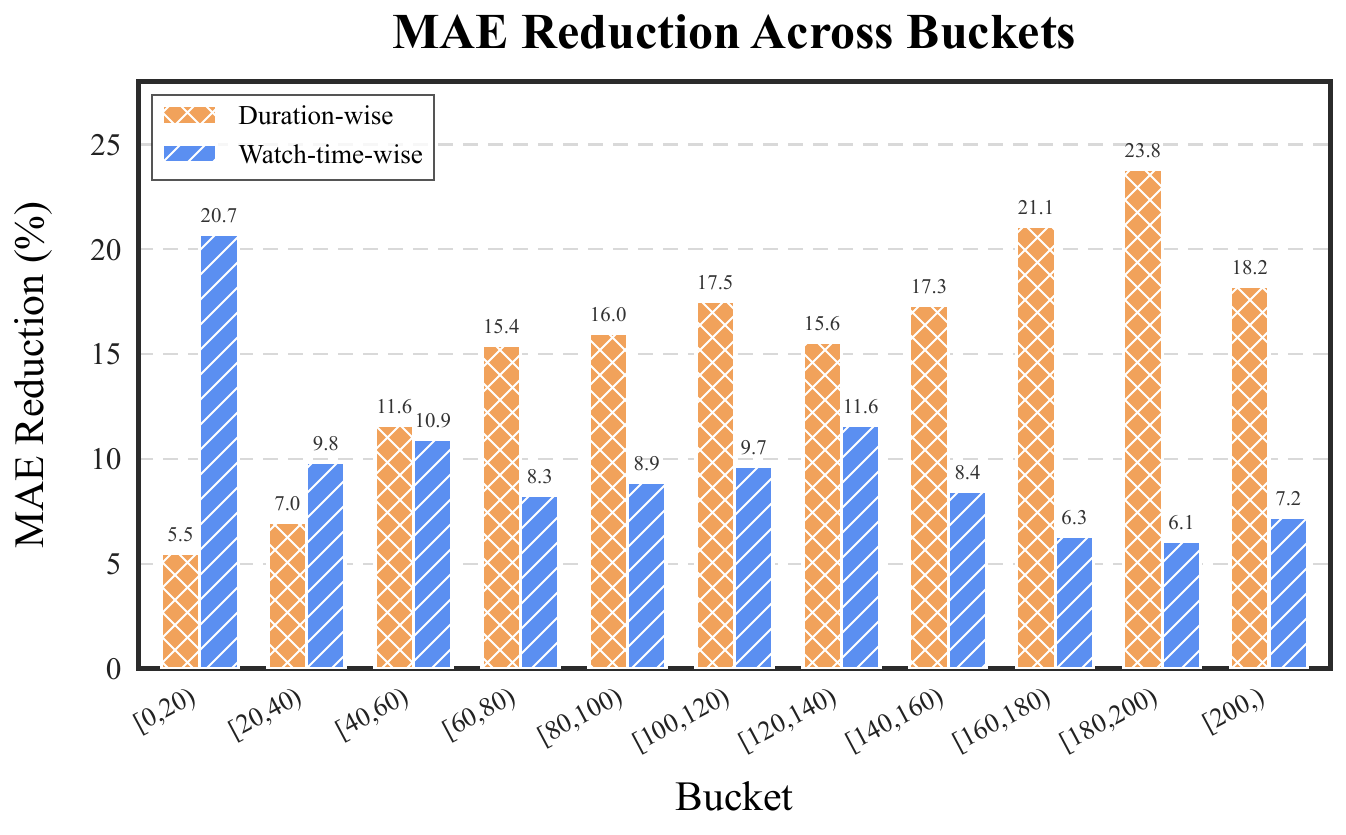}
    \caption{MAE reduction across duration-based serving regions and ex-post observed-watch-time regions.}
    \label{fig:03_mae_reduction_bucket}
\end{figure}

Figure~\ref{fig:03_mae_reduction_bucket} reports MAE reduction under two different partitions. Duration-wise buckets use an inference-time feature and evaluate performance across deployed routing regimes. Watch-time-wise buckets use the realized label and therefore provide only an ex-post localization of where errors are reduced.

DADF improves every displayed duration bucket: the reduction starts at 5.5\% and 7.0\% in the first two buckets, exceeds 15\% in most medium- and long-duration buckets, peaks at 23.8\% in $[180,200)$, and remains 18.2\% in $[200,+)$. In the label-space view, the largest reduction is 20.7\% for observed watch time in $[0,20)$, where Figure~\ref{fig:01_motivation}(b) also shows pronounced overestimation by the frozen predictor. The two marginal views establish that aggregate improvement is not produced by gains in only one duration regime and localize substantial correction to short-consumption samples. They do not identify duration as the cause of those errors.

\subsubsection{Long-Duration Tails}

\begin{figure}[t]
  \centering
  \includegraphics[width=\linewidth]{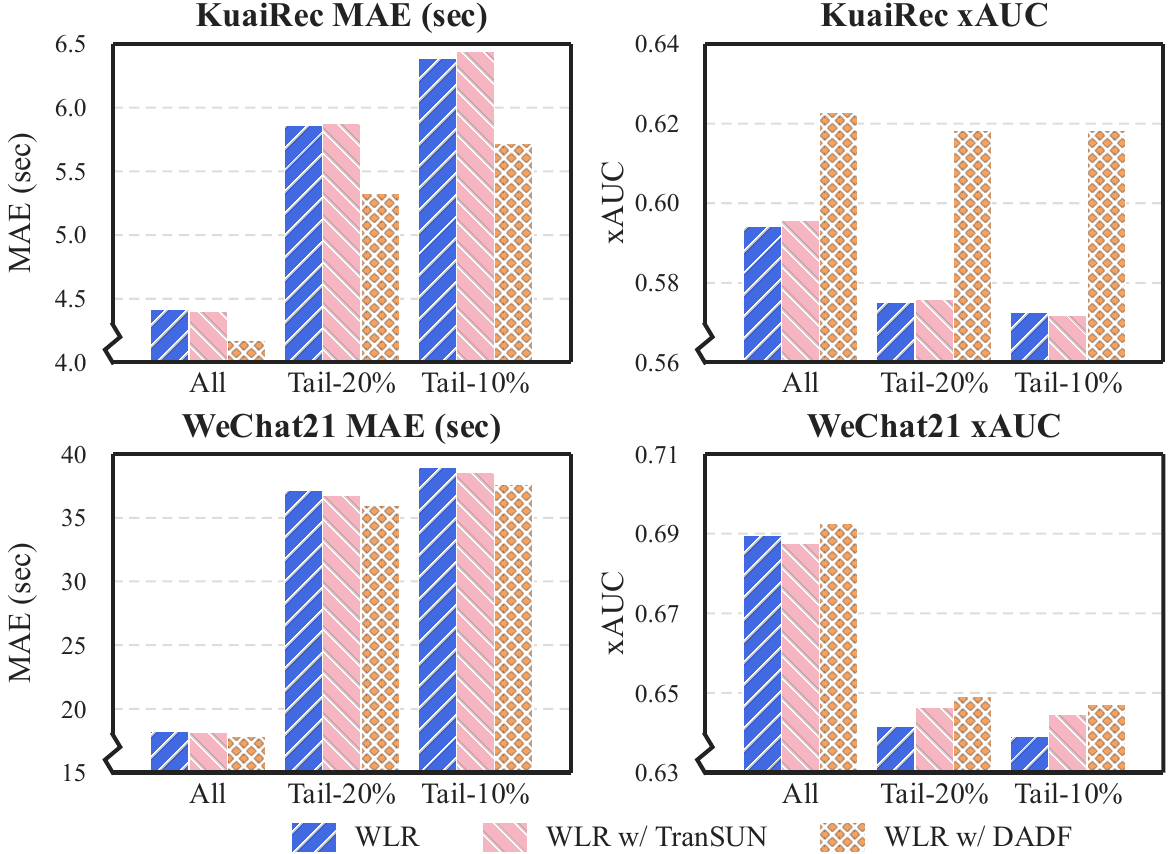}
  \caption{Tail-slice analysis on long-duration videos. Relative gains remain positive and increase as evaluation focuses on higher-duration slices.}
  \label{fig:05_tail_slice}
\end{figure}

Figure~\ref{fig:05_tail_slice} evaluates the top-duration Tail-20\% and Tail-10\% slices. On KuaiRec, MAE reduction increases from 5.48\% on all samples to 9.10\% and 10.43\%, while XAUC lift increases from 4.82\% to 7.52\% and 7.98\%. On WeChat21, MAE reduction increases from 2.26\% to 3.21\% and 3.35\%, and XAUC lift from 0.45\% to 1.21\% and 1.24\%. DADF therefore remains effective, and yields larger relative gains, in sparse long-duration slices. This is evidence of robustness under a shifted duration composition rather than evidence that duration itself generates the residual.

\subsubsection{Regime Granularity}

\begin{figure}[!h]
  \centering
  \includegraphics[width=\linewidth]{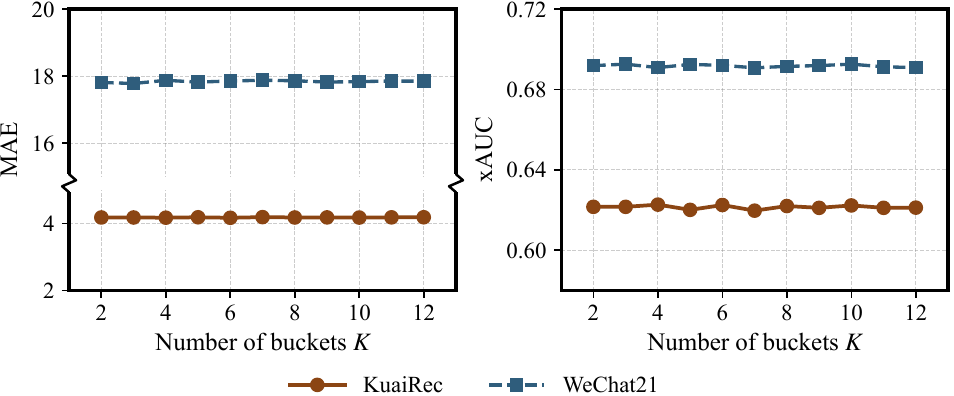}
  \caption{Sensitivity to the number of duration-indexed regimes $K$.}
  \label{fig:06_bucket_sensitivity}
\end{figure}

We vary $K=|\mathcal{G}|$ to test sensitivity to the granularity of $\pi(\cdot)$. Figure~\ref{fig:06_bucket_sensitivity} shows stable performance over a moderate range, including configurations coarser and finer than the production choice. Very fine partitions reduce per-regime support; production therefore uses $K=4$ as an engineering trade-off among validation performance, sample size, and serving complexity. This analysis demonstrates robustness to moderate choices of $K$, not optimality of the fixed boundaries. Adaptive regime construction under distribution drift remains future work.

\subsection{Production Evaluation and Deployment (RQ4)}

On production logs, DADF reduces overall MAE by 12.57\% and peak-hour MAE from 15.78 to 14.43. We then evaluate it in full ranking, rough ranking, and degraded serving, with the last also covering OneRec traffic~\cite{deng2025onerec}. Each experiment uses a seven-day A/A phase for traffic and metric validation followed by a seven-day A/B phase in which only the treatment enables DADF. The first-stage checkpoint, downstream objective, metric definitions, and candidate-generation logic remain unchanged.

Table~\ref{tab:02_online_ab} reports relative treatment--control lifts. Avg. Time Spent per Device, the primary metric, increases by 0.649\%, 0.235\%, and 0.199\% in the three stages, respectively; each primary lift is significant at the 95\% level~\cite{kohavi2013online,tang2010overlapping}. The stages differ in candidates, capacity, and serving constraints, so their lifts are neither pooled nor compared by magnitude. Gray acceptance regions are platform-provided zero-centered null regions, not confidence intervals around the reported lifts. After validation, DADF was deployed to 100\% of traffic in all three stages through the existing scalar watch-time interface.

\begin{table}[t]
\centering
\caption{
  Online A/B lifts. Bold marks the primary metric, and $^{*}$ denotes $p<0.05$.
  Completion Rate, for example, is the fraction of valid plays reaching completion.
  Arrows show the preferred direction. Short-view Rate is the fraction of valid plays
  with $0<y\leq3$ s, so lower is better. \textcolor{gray}{Gray subrows report the
  platform-provided 95\% null-hypothesis acceptance regions (ARs).}
}
\label{tab:02_online_ab}
\small
\renewcommand{\arraystretch}{1.08}
\setlength{\tabcolsep}{2.0pt}
\begin{tabular*}{\columnwidth}{@{\extracolsep{\fill}}c|ccc@{}}
\toprule
\multirow[c]{2}{*}{Metric} & Full & Rough & Degraded \\
                           & ranking & ranking & serving \\
\midrule
Total Time Spent (App)$\uparrow$       & +0.656\%$^{*}$ & +0.265\%$^{*}$ & +0.216\%$^{*}$ \\
\textcolor{gray}{\scriptsize 95\% AR} & \textcolor{gray}{\scriptsize[-0.14\%,0.14\%]} & \textcolor{gray}{\scriptsize[-0.17\%,0.17\%]} & \textcolor{gray}{\scriptsize[-0.12\%,0.12\%]} \\
Avg. Time Spent per Device$\uparrow$   & \textbf{+0.649\%}$^{*}$ & \textbf{+0.235\%}$^{*}$ & \textbf{+0.199\%}$^{*}$ \\
\textcolor{gray}{\scriptsize 95\% AR} & \textcolor{gray}{\scriptsize[-0.14\%,0.14\%]} & \textcolor{gray}{\scriptsize[-0.16\%,0.16\%]} & \textcolor{gray}{\scriptsize[-0.12\%,0.12\%]} \\
Completion Rate$\uparrow$               & +0.467\%$^{*}$ & +0.075\% & +2.210\%$^{*}$ \\
\textcolor{gray}{\scriptsize 95\% AR} & \textcolor{gray}{\scriptsize[-0.11\%,0.11\%]} & \textcolor{gray}{\scriptsize[-0.13\%,0.13\%]} & \textcolor{gray}{\scriptsize[-0.10\%,0.10\%]} \\
Effective-view Rate$\uparrow$           & +0.650\%$^{*}$ & +0.250\%$^{*}$ & +1.173\%$^{*}$ \\
\textcolor{gray}{\scriptsize 95\% AR} & \textcolor{gray}{\scriptsize[-0.11\%,0.11\%]} & \textcolor{gray}{\scriptsize[-0.13\%,0.13\%]} & \textcolor{gray}{\scriptsize[-0.10\%,0.10\%]} \\
Long-view Rate$\uparrow$                & +0.675\%$^{*}$ & +0.360\%$^{*}$ & +1.607\%$^{*}$ \\
\textcolor{gray}{\scriptsize 95\% AR} & \textcolor{gray}{\scriptsize[-0.13\%,0.13\%]} & \textcolor{gray}{\scriptsize[-0.15\%,0.15\%]} & \textcolor{gray}{\scriptsize[-0.11\%,0.11\%]} \\
Short-view Rate$\downarrow$             & -0.477\%$^{*}$ & -0.109\% & -0.515\%$^{*}$ \\
\textcolor{gray}{\scriptsize 95\% AR} & \textcolor{gray}{\scriptsize[-0.10\%,0.10\%]} & \textcolor{gray}{\scriptsize[-0.12\%,0.12\%]} & \textcolor{gray}{\scriptsize[-0.09\%,0.09\%]} \\
\bottomrule
\end{tabular*}
\end{table}

\section{Conclusion}

We studied second-stage correction for frozen watch-time predictors that are
well balanced in aggregate and approximately calibrated with respect to their
own scores, yet retain conditional residual structure that can be predicted
from inference-time signals. DADF estimates a multiplicative correction from
the full serving representation, stabilizes its heavy-tailed target through
regime-specific transformations, uses video duration only to index
heterogeneous correction regimes, and incorporates auxiliary engagement
representations. Across seven backbones on two public datasets, DADF reduces
MAE by 4.33\% and improves XAUC by 4.01\% on average. In production, it reduces
overall MAE by 12.57\% and yields significant Avg. Time Spent per Device lifts
of 0.649\%, 0.235\%, and 0.199\% across three serving stages, all of which were
subsequently deployed at 100\% traffic.

These results support the practical value of correcting the predictable
component of conditional residuals without changing the first-stage prediction
interface. They do not imply that all label-space mean shrinkage is removable
or that duration causes the residual pattern. Future work will study adaptive
regime construction under distribution drift and correction objectives beyond
pointwise regression losses.

\clearpage
\bibliographystyle{ACM-Reference-Format}
\bibliography{references}

\appendix
\section{Theoretical Details}

\subsection{Long-Tailedness Inheritance of the Multiplicative Correction Target}
\label{app:long_tail_factor}

We first formalize why dividing watch time by a fixed positive first-stage
prediction does not, by itself, eliminate a long tail. Let \(S\) denote the
inference-time signals defined in Section~\ref{sec:problem}, and
let
\begin{equation}
  c(S)=\max(\operatorname{sg}(\hat{Y}_0),\epsilon)>0,
  \qquad
  B=\frac{Y}{c(S)}.
\end{equation}
For a fixed serving state \(S=s\), define the conditional survival function
\(\bar F_Y(t\mid s)=\mathbb{P}(Y>t\mid S=s)\). Suppose that
\(Y\mid S=s\) belongs to the long-tailed class \(\mathcal{L}\), i.e.,
\begin{equation}
  \lim_{t\rightarrow\infty}
  \frac{\bar F_Y(t+a\mid s)}{\bar F_Y(t\mid s)}=1
\end{equation}
for every fixed \(a\in\mathbb{R}\) for which the survival function is
eventually positive. Since \(c(s)\) is a fixed positive constant under the
conditioning,
\begin{equation}
  \bar F_B(t\mid s)
  =
  \mathbb{P}\!\left(Y>t c(s)\mid S=s\right)
  =
  \bar F_Y(t c(s)\mid s).
\end{equation}
Therefore, for any fixed shift \(a\),
\begin{equation}
\begin{aligned}
  \frac{\bar F_B(t+a\mid s)}{\bar F_B(t\mid s)}
  &=
  \frac{\bar F_Y((t+a)c(s)\mid s)}
       {\bar F_Y(tc(s)\mid s)} \\
  &=
  \frac{\bar F_Y(u+a c(s)\mid s)}
       {\bar F_Y(u\mid s)}
  \longrightarrow 1,
\end{aligned}
\end{equation}
where \(u=tc(s)\rightarrow\infty\). Hence \(B\mid S=s\in\mathcal{L}\).
The result is conditional and does not assert that every empirical
correction-target distribution is long-tailed. It shows that ratio
construction alone provides no general tail-removal guarantee, motivating
the target stabilization described in Section~\ref{sec:dynamic_distribution}.

\begin{figure*}[!t]
  \centering
  \includegraphics[width=\linewidth]{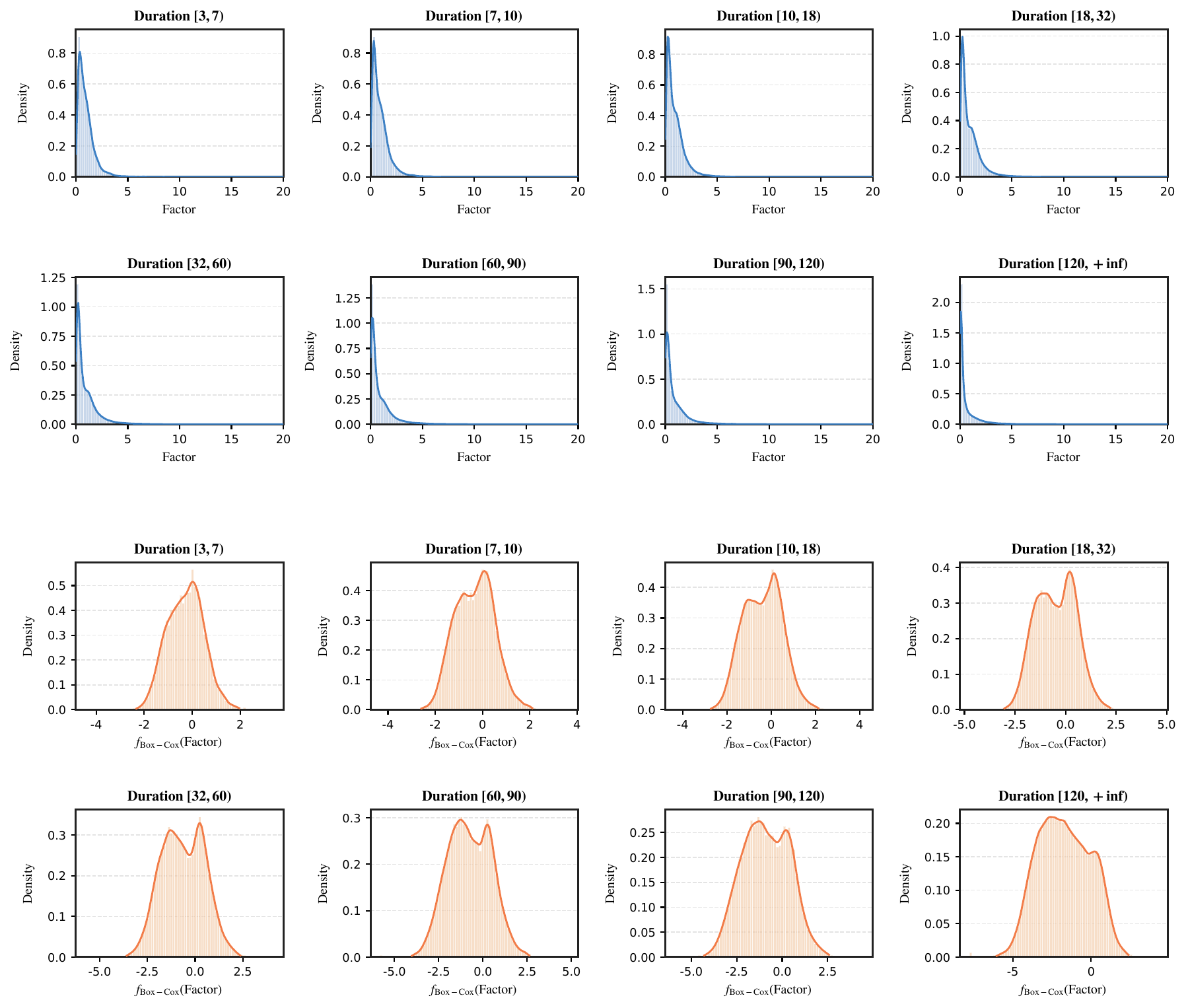}
  \caption{Distribution comparison of the raw multiplicative correction target
  (top) and its group-specific Box--Cox transformation (bottom) on Kwai. The
  raw target is right-skewed in each duration-indexed regime, while the
  transformed target is more compact. This is a descriptive distributional
  diagnostic rather than evidence that duration causes the residual.}
  \label{fig:07_factor_distribution_compare}
\end{figure*}

\subsection{Risk Decomposition for Predictable Correction}
\label{app:predictable_risk_proof}

We give the squared-loss decomposition used to define the correctable
component of the residual. Let \(S\) contain only inference-time signals and
let \(f(S)\) be any second-stage predictor. The population risk is
\begin{equation}
  \mathcal{R}(f)=\mathbb{E}\!\left[(Y-f(S))^2\right].
\end{equation}
The Bayes predictor is \(f^\star(S)=\mathbb{E}[Y\mid S]\). Writing
\(Y-f(S)=(Y-f^\star(S))+(f^\star(S)-f(S))\) and conditioning the cross term
on \(S\) gives
\begin{equation}
  \mathcal{R}(f)
  =
  \mathbb{E}\!\left[\operatorname{Var}(Y\mid S)\right]
  +
  \mathbb{E}\!\left[(f^\star(S)-f(S))^2\right].
  \label{eq:appendix_risk_decomposition}
\end{equation}
Applying Equation~\ref{eq:appendix_risk_decomposition} to the frozen
first-stage prediction \(f(S)=\hat{Y}_0\) yields
\begin{equation}
  \mathcal{R}(\hat{Y}_0)-\mathcal{R}(f^\star)
  =
  \mathbb{E}\!\left[
    \bigl(\mathbb{E}[Y-\hat{Y}_0\mid S]\bigr)^2
  \right].
\end{equation}
Thus the maximum population MSE reduction available to an inference-time
corrector is exactly the squared magnitude of the predictable conditional
residual. The remaining conditional variance is irreducible under the
available signals. This oracle identity does not guarantee that a finite
DADF model will attain the Bayes risk, nor does it directly imply
improvements in MAE or XAUC; those claims are evaluated empirically.

\subsection{What Duration-Indexed Routing Can and Cannot Guarantee}
\label{app:group_risk_proof}

Let \(R\) denote the non-routing inputs to a correction model, \(G=\pi(d)\)
the duration-indexed regime, and \(U\) the transformed correction target.
For a fixed expert architecture \(q_{\theta}(R,G)\), define the shared class
\(\mathcal{F}_0=\{q_{\theta}:\theta\in\Theta\}\) and the hard-routed class
\begin{equation}
  \mathcal{F}_{G}
  =
  \left\{
    \sum_{g=1}^{K}\mathbf{1}(G=g)q_{\theta_g}(R,G):
    \theta_1,\ldots,\theta_K\in\Theta
  \right\}.
\end{equation}
If every expert has the same functional form as the shared model, then
\(\mathcal{F}_0\subseteq\mathcal{F}_{G}\): setting
\(\theta_1=\cdots=\theta_K\) recovers the shared predictor. Consequently,
\begin{equation}
  \inf_{q\in\mathcal{F}_{G}}
  \mathbb{E}\!\left[(U-q(R,G))^2\right]
  \le
  \inf_{q\in\mathcal{F}_0}
  \mathbb{E}\!\left[(U-q(R,G))^2\right].
\end{equation}
This is a function-class inclusion result, not a claim of guaranteed
generalization improvement. The inequality can be strict only when the
partitioned class represents useful target heterogeneity that the shared
class cannot represent equally well. With finite data, additional experts
can also increase estimation error.

Moreover, when duration is already included in \(R\), the regime \(G\) is a
deterministic function of the existing input and adds no new oracle
information. In that common setting, hard routing changes the inductive bias
and parameter specialization rather than the information set. The
effectiveness of duration-indexed routing must therefore be established by
the matched-protocol ablation in Section~\ref{sec:experiments}, and should not
be interpreted as evidence that duration causes the residual.

\section{Additional Experimental Results}

\subsection{Correction-Target Stabilization Across Duration-Indexed Regimes}

Figure~\ref{fig:07_factor_distribution_compare} compares the raw
multiplicative target with its group-specific Box--Cox transformation. The
raw distributions are right-skewed across all displayed regimes, while their
dispersion and fitted shapes vary with the duration index. After
transformation, the targets are more compact within each regime. These
descriptive results motivate regime-specific target stabilization; the
performance contribution of the transformation is evaluated by the
``w/o Dist'' ablation in Table~\ref{tab:03_ablation}.

\begin{figure}[t]
  \centering
  \includegraphics[width=\linewidth]{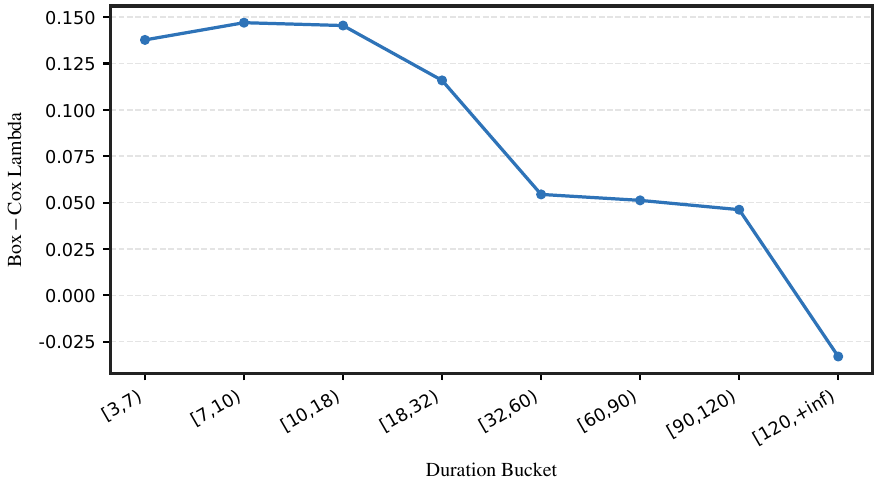}
  \caption{Fitted group-specific Box--Cox parameters \(\lambda_g\) across
  duration-indexed regimes on Kwai. The variation is descriptive evidence of
  heterogeneous target geometry, not a causal effect of duration.}
  \label{fig:08_boxcox_lambda}
\end{figure}

Figure~\ref{fig:08_boxcox_lambda} reports the fitted Box--Cox parameter
\(\lambda_g\) for each duration-indexed regime. The fitted values are not
constant and generally decrease for longer-duration regimes, indicating that
a single global transformation would impose the same target geometry on
empirically different marginal distributions. This observation supports the
design hypothesis but does not, on its own, establish that duration supplies
incremental predictive information or that group-specific transformation is
optimal.

\end{document}